\documentclass[12pt]{article}
\usepackage[latin9]{inputenc}
\usepackage{geometry}
\usepackage{amsmath}
\usepackage{amssymb}
\usepackage{graphicx}
\usepackage{esint}

\makeatletter
\newcommand{\lyxaddress}[1]{
\par {\raggedright #1
\vspace{1.4em}
\noindent\par}
}

\@ifundefined{date}{}{\date{}}

\usepackage{epstopdf}
\usepackage{cite}

\makeatother

\begin{document}

\title{New approach to $K^{0}-\bar{K^{0}}$ mixing}

\author{N. F. Nasrallah}
\maketitle

\lyxaddress{}

\lyxaddress{\begin{center}
Faculty of Science, Lebanese University. Tripoli 1300, Lebanon 
\par\end{center}}
\begin{abstract}
A new QCD calculation of the $B$ parameter of $K^{0}-\bar{K^{0}}$
mixing is presented. It makes use of polynomial kernels in dispersion
integrals in order to practically eliminate the contributions of the
unknown pseudoscalar strange continuum. This approach avoids the arbitrariness
and instability inherent to the Borel exponential kernels used in
previous sum rules calculations. A simultaneous calculation of the
mixed quark gluon condensate $\langle g\bar{q}\sigma_{\mu\nu}\frac{\lambda^{a}}{2}\sigma_{\mu\nu}^{a}q\rangle$
which enters in the expression for $B$ is presented. Finally the
K-meson decay constant $f_{K}$ is calculated to five loops.\\
\\
Pacs Numbers: 14.40.Df, 13.25.Es, 11.55.Hx
\end{abstract}

\section{Introduction}

In the Standard Model (SM) the mixing of the two eigenstates of strangeness
\cite{SM} is predicted as a higher order process \cite{GL} which
contributes to the $K_{L}-K_{S}$ mass difference through the so-called
$\Delta S=2$ box diagram.

The $K_{L}-K_{S}$ mass difference $\Delta m$ is a sum of a long
distance dispersive contribution $\Delta m_{L}$ and a short distance
one $\Delta m_{S}$ proportioned to the matrix element 
\begin{equation}
\langle\bar{K^{0}}(p')|\theta_{\Delta S=2}|K^{0}(p)\rangle
\end{equation}
with 
\begin{equation}
\theta_{\Delta S=2}=(\bar{s}\gamma_{\mu}(1-\gamma_{s})d)(\bar{s}\gamma_{\mu}(1-\gamma_{s})d)
\end{equation}
Neglecting anomalous dimension factors the parameter $B$ is defined
as 
\begin{equation}
\langle\bar{K^{0}}(p')|\theta_{\Delta S=2}|\bar{K^{0}}(p)\rangle=\frac{16}{3}Bf_{K}^{2}(p.p')
\end{equation}

And $B=1$ in the vacuum saturation approximation.

More sophisticated calculations of $B$ followed using quark and bag
models, lattice calculations and $QCD$ sum rule techniques \cite{PV,Bil,RY,D,P,cheter,Bru,Aoki}.
Unfortunately no single value for $B$ bas emerged.

I present here a new approach. Starting with a 3-pt function involving
two pseudoscalar currents in addition to the $\Delta S=2$ four quark
operator 
\begin{equation}
A(p,p').(p.p')=i^{2}\iint dx\:dy\:e^{ipx-ip'y}\:\langle0|Tj_{5}(x)\:\theta_{\Delta S=2}(0)\:j_{5}(y)|0\rangle
\end{equation}

Dispersion relations for this quantity will be written and intermediate
states inserted. The K-meson poles carry the sought for information
in addition there is the contribution of the strange pseudoscalar
continuum of which not much is known except that it is dominated by
two radial excitations of the K, K (1460) and K (1830). In order to
damp the unknown contribution of the continuum Borel (Laplace) \cite{SVZ}
transforms of correlators have been used in which case the damping
is provided by an exponential kernel $e^{-t/M^{2}}$. If the parameter
$M^{2}$, the squared of the Borel mass, is small the damping is good
but the contribution of the unknown higher order condensates increase
rapidly. If $M^{2}$ increases the contribution of the unknown condensates
decreases but the damping in the resonance region worsens. An intermediate
value of $M^{2}$ has to be chosen. Because $M^{2}$ is an unphysical
parameter the results should be independent of it in a relatively
broad window which is not always the case. The choice of the parameter
which signals the onset of perturbative $QCD$ is another source of
uncertainty.

In this work I shall use polynomial kernels in order to eliminate
the contributions of the unknown continuum. The coefficients of these
polynomials are chosen to make the roots coincide with the masses
of the radial excitation of the K and involve none of the arbitrariness
and instability inherent to the use of exponential kernels.

$A(p,p')^{QCD}$ is made of a factorisable $(f)$ and a non-factorisable
$(nf)$ part. The $nf$ part is dominated by the non-perturbative
quark gluon mixed condensate. 
\begin{equation}
g\:\langle\bar{q}\:\sigma_{\mu\nu}\frac{\lambda^{2}}{2}G_{\mu\nu}^{(a)}\:q\rangle\equiv m_{0}^{2}\:\langle\bar{q}q\rangle
\end{equation}

Estimates of $m_{0}^{2}$ in the literature range from 0.3 $GeV^{2}$
to 1.2 $GeV^{2}$. The method used here allows for an independent
determination of $m_{0}^{2}$ which will be used. As a final application
the value the K decay constant is calculated to 5-loops.

\section{Calculation of the $B$ parameter}

Consider the 3-point function 
\begin{equation}
A(p,p').(p.p')=i^{2}\iint dx\:dy\:e^{ipx-ip'y}\:\langle0|Tj_{5}(x)\:\theta_{\Delta S=2}(0)\:j_{5}(y)|0\rangle
\end{equation}
where $j_{5}(x)=\bar{d}(x)i\,\gamma_{5}\,s(x)$ is the pseudoscolar
current. The amplitude $A(t=p^{2},\,t'=p'^{2},\,p.p')$ will be studied
at fixed $p.p'$ and will be denoted by $A(t,t')$.

$A(t,t')$ possesses a double pole, two single poles and cuts on the
real $t,t'$ axes which extend from $th={(m_{K}+2m_{\pi})}^{2}$ to
infinity stemming from the strange pseudoscalar intermediate states.

\begin{equation}
A(t,t')(p.p')=\frac{2f_{K}^{2}m_{K}^{4}\:\langle\:K^{0}|\theta_{\Delta S=2}|\bar{K^{0}}\:\rangle}{(m_{s}+m_{d})^{2}(t-m_{K}^{2})(t'-m_{K}^{2})}+\frac{\Phi(t)}{(t'-m_{K}^{2})}+\frac{\Phi(t')}{(t-m_{K}^{2})}+...
\end{equation}

Consider now the double integral in the complex $t$ and $t'$ planes

\[
\frac{1}{(2\pi i)^{2}}\int_{c}\int_{c'}dt\:dt'P(t)P(t')A(t,t')(p.p')
\]

where $c$ and $c'$ are the contours shown on fig.1, $f_{K}$ is
the $K$ decay constant $(f_{K}=.114\:GeV)$ and $P(t)$ is a so far
arbitrary entire function.

Because $\Phi(t),\Phi(t')$ have no singularities inside the contours
of integration the single poles do not contribute to the double integral
and we are left with

\begin{equation}
\begin{aligned} & \frac{2f_{K}^{2}m_{K}^{4}}{(m_{s}+m_{d})^{2}}\langle\:K^{0}|\theta_{\Delta S=2}|\bar{K}^{0}\:\rangle P^{2}(m_{K}^{2})=\\
 & \frac{1}{(2\pi i)^{2}}\int_{th}^{R}dt\:P(t)\int dt'\:P(t')\:Disc_{t,t'}A(t,t')(p.p')\\
+ & \frac{1}{(2\pi i)^{2}}\oint dt\:P(t)\oint dt'\:P(t')\:A(t,t')(p.p')
\end{aligned}
\label{eq:8}
\end{equation}

The first integral on the r.h.s of eq. (\ref{eq:8}) represents the
contribution of the pseudoscalar strange continuum. $P(t)$ is now
chosen to be a second order polynomial whose roots coincide with the
masses squared of the radial excitations of the $K,\:K(1460)$ and
$K(1870)$ 
\begin{equation}
P(t)=1-.768\:GeV^{-2}t\:+\:.14\:GeV^{-4}t^{2}
\end{equation}

$P(t)$ and $P(t')$ essentially eliminate the contributions of the
continuum. On the circle of large radius $R\quad A(t,t')$ can safely
be replaced by $A^{QCD}(t,t')$. 

So that, using $\langle K^{0}\:|\:\theta_{\Delta S=2}\:|\:\bar{K^{0}}\rangle=\frac{16}{3}f_{K}^{2}\:B(p.p')$
gives

\begin{equation}
\frac{32}{3}\dfrac{f_{K}^{4}m_{K}^{4}}{(m_{s}+m_{d})^{2}}P^{2}(m_{K}^{2})B=\dfrac{1}{(2\pi i)^{2}}\oint\oint dt\:dt'P(t)P(t')A^{QCD}(t,t')
\end{equation}

$A^{QCD}(t,t')$ is the sum of a factorisable and a non-factorisable
part \cite{RY,Bil} 
\begin{equation}
A^{QCD}=A_{f}^{QCD}+A_{nf}^{QCD}
\end{equation}

where

\begin{equation}
A_{f}^{QCD}=\frac{8}{3}\Pi_{5}(t)\:\Pi_{5}(t')
\end{equation}

\begin{equation}
\Pi_{5}(t)=-\frac{3}{8\pi^{2}}m_{s}ln(-t)+\dfrac{\langle\bar{d}d+\bar{s}s\rangle}{t}-\dfrac{m_{s}\langle a_{s}GG\rangle}{8}\frac{1}{t^{2}}+\frac{0}{t^{3}}+...
\end{equation}

and

\begin{equation}
\begin{aligned}A_{nf}^{QCD} & =\frac{2}{3}m_{0}^{2}\langle\bar{q}q\rangle^{2}\:(\dfrac{1}{t^{2}t'}+\dfrac{1}{t'^{2}t})\\
 & +\frac{1}{4\pi^{2}}m_{0}^{2}\langle m_{s}\bar{q}q\rangle\:\frac{1}{tt'}\:(ln(-\frac{t}{\mu^{2}})+ln(-\frac{t'}{\mu^{2}}))\\
 & -[\frac{4\pi^{2}}{9}\langle\bar{q}q\rangle^{2}\langle a_{s}GG\rangle+\frac{13}{288}m_{0}^{4}\langle\bar{q}q\rangle^{2}]\:\dfrac{1}{t^{2}t'^{2}}+...
\end{aligned}
\end{equation}

The expressions for $B_{f}$ and $B_{nf}$ follow: 
\begin{equation}
\dfrac{4f_{K}^{4}m_{K}^{4}}{(m_{s}+m_{d})^{2}}P^{2}(m_{K}^{2})B_{f}=I^{2}
\end{equation}

where 
\begin{equation}
\begin{aligned}I= & \frac{1}{2\pi i}\oint dtP(t)\Pi_{5}(t)\\
= & -\frac{3m_{s}}{8\pi^{2}}\frac{1}{2\pi i}\oint dt\:P(t)ln(-t)+\langle\bar{d}d+\bar{s}s\rangle+\frac{a_{1}m_{s}}{8}\langle a_{s}GG\rangle
\end{aligned}
\end{equation}

Because $ln(-t)$ posses a cut on the positive $t-axis$ which starts
at the origin, the integral over the circle in the equation above
can be transformed in an integral over the real axis so that

\begin{equation}
I=-\frac{3m_{s}}{8\pi^{2}}\int_{0}^{R}dt\:P(t)+\langle\bar{d}d+\bar{s}s\rangle+\frac{a_{1}m_{s}}{8}\langle a_{s}GG\rangle\label{eq:17}
\end{equation}

The choice of $R$ is determined by stability considerations. It should
not be too small as this would invalidate the operator product expansion
on the circle, nor should it be too large because $P(t)$ would start
enhancing the contribution of the continuum instead of suppressing
it. We seek an intermediate range of $R$ for which the integral (\ref{eq:17})
is stable.

As discussed above, our choice for $P(t)$ is $P(t)=1-.768t+.140t^{2}$
which vanishes at the radial excitations of the $K$ and is very small
in a broad interval around them. The integral $i(R)=\int_{0}^{R}dt\:P(t)$
is seen to be stable for $2\:GeV^{2}\lesssim R\lesssim4\:GeV^{2}$,
$i(R)\approx.83\:GeV^{2}$ as shown in fig.2. Then

\begin{equation}
\begin{aligned}-(m_{s}+m_{d})I= & -(m_{s}+m_{d})\langle\bar{d}d+\bar{s}s\rangle\\
+ & \frac{3}{8\pi^{2}}m_{s}(m_{s}+m_{d})\:i(R)-\frac{a_{1}m_{s}}{8}(m_{s}+m_{d})\langle a_{s}GG\rangle
\end{aligned}
\end{equation}

Turn now to the contribution of the $nf$ part. Similar manipulations
lead to 
\begin{equation}
\begin{aligned} & \frac{4f_{K}^{4}m_{K}^{4}}{(m_{s}+m_{d})^{2}}\:P^{2}(m_{K}^{2})\:B_{nf}=\\
 & -\dfrac{1}{2}a_{1}m_{0}^{2}\langle\bar{q}q\rangle^{2}+\frac{3m_{0}^{2}}{16\pi^{2}}\langle m_{s}\bar{q}q\rangle\:ln\frac{R}{\mu^{2}}\\
 & -\frac{3}{8}[\:\frac{4\pi^{2}}{9}\langle\bar{q}q\rangle^{2}\langle a_{s}GG\rangle+\frac{13}{288}m_{0}^{4}\langle\bar{q}q\rangle^{2}]\:a_{1}^{2}
\end{aligned}
\label{eq:19}
\end{equation}

Values of $m_{0}^{2}$, which parametrizes the quark-gluon mixed condensate
vary over a large range in the literature. The method presented here
allows an independent evaluation of this quantity:

The integral 
\begin{equation}
\dfrac{1}{(2\pi i)^{2}}\int_{c}^ {}P(t)(t-m_{K}^{2})\int_{c}^ {}dt'P(t')(t'-m_{K}^{2})\:A^{QCD}(t,t')=I'_{f}+I'_{nf}=0\label{eq:20}
\end{equation}

Vanishes because the singularities inside $C,C'$ have been removed.
\begin{equation}
I'_{f}=\frac{8}{3}I'^{2}
\end{equation}

\begin{equation}
\begin{aligned} & I'=\frac{1}{2\pi i}\oint dt\:P(t)(t-m_{K}^{2})\{-\frac{3m_{s}}{8\pi^{2}}\:ln-t+\frac{\langle\bar{d}d+\bar{s}s\rangle}{t}-\frac{m_{s}}{8}\langle a_{s}GG\rangle\frac{1}{t^{2}}\}\\
 & =-m_{K}^{2}\langle\bar{d}d+\bar{s}s\rangle-\frac{3m_{s}}{8\pi^{2}}\:i'(R)-\frac{m_{s}}{8}\langle a_{s}GG\rangle(1+a_{1}m_{K}^{2})
\end{aligned}
\end{equation}

where

\begin{equation}
i'(R)=\int_{0}^{R}dt\:P(t)(t-m_{K}^{2})
\end{equation}

The n.f contribution is

\begin{equation}
\begin{aligned}I'_{nf}= & -\frac{4}{3}m_{0}^{2}{\langle\bar{q}q\rangle}^{2}\:m_{K}^{2}(1+a_{1}m_{K}^{2})\\
 & -\frac{m_{s}}{2\pi^{2}}m_{0}^{2}\langle\bar{q}q\rangle\:m_{K}^{2}[i''(R)-m_{K}^{2}ln\frac{R}{\mu^{2}}]\\
 & -[\frac{4\pi^{2}}{9}\langle\bar{q}q\rangle^{2}\langle a_{s}GG\rangle+\frac{13}{288}m_{0}^{4}\langle\bar{q}q\rangle^{2}]\:(1+a_{1}m_{\alpha}^{2})^{2}
\end{aligned}
\end{equation}

where

\begin{equation}
i''(R)=\int_{0}^{R}dt\:[1+a_{1}m_{K}^{2}-(a_{1}-a_{2}\:m_{K}^{2})t-a_{2}t^{2}]\label{eq:25}
\end{equation}

Eqs (\ref{eq:20}, \ref{eq:25}) determine $m_{0}^{2}$ which in turn
yields $B_{nf}$ when inserted in eq.(\ref{eq:19}).

The condensate $\langle\bar{d}d+\bar{s}s\rangle$ dominates our equations.
This quantity was determined in \cite{DNS} from a study of the 2-point
correlator

\[
\Psi_{5}(t)=i\int dx\:e^{iqx}\:\langle0|T\partial_{\mu}A_{\mu}^{ds}(x)\partial_{\nu}A_{\nu}^{ds}(0)|0\rangle
\]

with the result 
\begin{equation}
\begin{aligned}\Psi_{5}(0)= & -(m_{s}+m_{d})\langle\bar{d}d+\bar{s}s\rangle\\
= & 2f_{K}^{2}m_{K}^{2}P(m_{K}^{2})+\delta_{5}
\end{aligned}
\end{equation}

and it was obtained

\begin{equation}
\delta_{5}\approx\frac{1}{2\pi i}\oint\frac{dt}{t}\:P(t)\:\Psi_{(t)}^{QCD}=.0014\:GeV^{4}
\end{equation}
or 
\begin{equation}
\Psi_{5}(0)=(.39\pm.03).10^{-2}\:GeV^{4}
\end{equation}

This, with $(m_{s}+m_{d})=(108\pm8)\:MeV$, 

yield finally

\begin{equation}
\begin{aligned}m_{0}^{2}=.90\:GeV^{2}\\
B=.55\pm.08
\end{aligned}
\end{equation}

\section{$f_{K}$ to five loops}

The method is further used in the calculation of the kaon decay constant
$f_{K}$. Start with the correlator 
\[
\begin{aligned}\Pi_{\mu\nu}(t=q^{2})= & i\int dx\:e^{iqx}\langle0|T\:A_{\mu}(x)A_{\nu}(0)|0\rangle\\
= & (q_{\mu}q_{\nu}-q_{\mu\nu}\:q^{2})\Pi^{(1)}(t)+q_{\mu}q_{\nu}\Pi^{(0)}(t)
\end{aligned}
\]

$\Pi(t)=\Pi^{(0+1)}(t)$ And consider

\begin{equation}
\begin{aligned}\int_{c}dt\:P(t)\Pi(t)= & 2f_{K}^{2}P(m_{K}^{2})\\
= & \frac{1}{\pi}\int_{0}^{R}dt\:P(t)\:Im\:\Pi(t)+\frac{1}{2\pi i}\oint dt\:P(t)\:\Pi^{QCD}(t)
\end{aligned}
\end{equation}

As before the polynomial $P(t)$ is chosen in order to eliminate the
contribution of the integral on the cut. We have now to take into
account the axial-vector resonances in addition to the pseudoscalar
ones. That is $K_{1}(1273)$ and $K_{1}(1402)$ in addition to $K(1460)$
and $K(1830)$. The choice 
\begin{equation}
P(t)=1-1.42t+.648t^{2}-.093t^{3}
\end{equation}
with the coefficients in appropriate powers of $GeV$ achieves the
purpose of eliminating the contribution of the continuum. Here 
\begin{equation}
\Pi^{QCD}(t)=\Pi_{pert}(t)+\frac{c_{1}}{t}+\frac{c_{2}}{t^{2}}+\frac{c_{3}}{t^{3}}+...
\end{equation}

\begin{equation}
4\pi\:Im\:\Pi_{pert}(t)=1+a_{s}(r)+a_{s}^{2}(r)\:l_{2}(t,r)+a_{s}^{3}(r)\:l_{3}(t,r)+a_{s}^{4}(r)\:l_{4}(t,r)
\end{equation}

The $l_{i}(t,r)$ and the strong coupling constant $a_{s}(r)$ are
known to 5-loop order \cite{Bai,cheter} and the non-perturbative
condensates are given by \cite{Bru}.

\begin{equation}
\begin{aligned}c_{1}= & +\frac{3m_{s}^{2}}{4\pi^{2}}(1+\frac{7}{3}a_{s})\\
c_{2}= & \frac{1}{12}(1-\frac{11}{18}a_{s})\langle a_{s}GG\rangle+(1-\frac{a_{s}}{3})\langle m_{s}\bar{s}s\rangle\\
c_{3}= & -a_{s}\frac{32\pi^{2}}{9}[\:\langle\bar{q}q\rangle\:\langle\bar{s}s\rangle-\frac{1}{9}\langle qq\rangle^{2}-\frac{1}{9}\langle\bar{s}s\rangle\:]
\end{aligned}
\end{equation}

The integral of $\Pi_{pert}(t)$ over the circle is transformed into
an integral over the cut once again and finally

\begin{equation}
2f_{K}^{2}P(m_{K}^{2})=\frac{1}{\pi}\int_{0}^{R}dt\:P(t)\:Im\:\Pi_{pert}(t)+c_{1}-a_{1}c_{2}-a_{2}c^{3}
\end{equation}

The integral is stable for $1\lesssim r\lesssim3.5\:GeV^{2}$ and
the non perturbative contribution are small. The final result is 
\begin{equation}
f_{K}=(.110\pm.003)\:GeV
\end{equation}

The error is obtained by varying the renormalization scale $\mu$
which enters in the $QCD$ expressions $\Pi_{pert}$. 
\begin{figure}
\centering{}\includegraphics{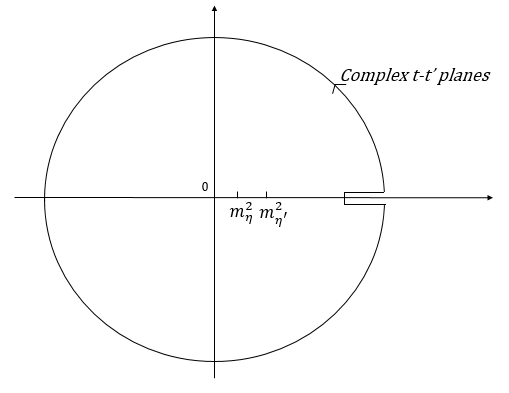} \caption{The contours of integration c,c'.}
\end{figure}

\begin{figure}
\centering{}\includegraphics{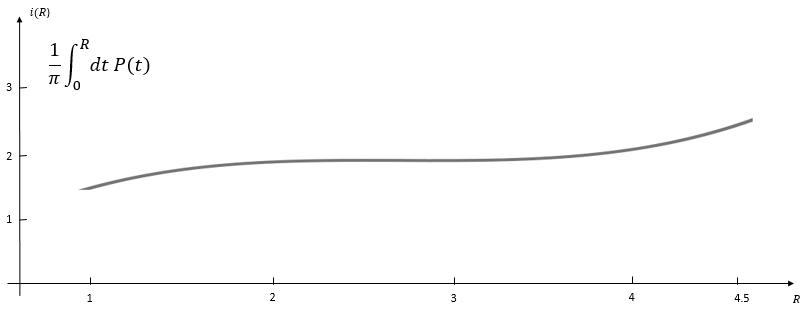} \caption{The variation of $i(R)=\int_{0}^{R}dtP(t)$ as a function of $R$
in $Gev$.}
\end{figure}
\newpage{}

\end{document}